\documentclass{ifacconf}

\usepackage{graphicx}      
\usepackage{natbib}        

\usepackage{amsmath,amssymb,amsfonts}
\usepackage{siunitx}
\usepackage{bm}
\usepackage{graphicx}
\usepackage{xcolor}
\usepackage{url}
\usepackage{algorithm}
\usepackage{algpseudocode}
\graphicspath{{pics/}}

\usepackage{tikz}
\usetikzlibrary{patterns}

\newcommand{\D}[1]{\Delta{#1}}
\newcommand{\mat}[1]{\ensuremath{\begin{bmatrix} #1 \end{bmatrix}}}
\newcommand{\vc}[1]{\ensuremath{\begin{pmatrix} #1 \end{pmatrix}}}
\newcommand{\ol}[1]{\overline{#1}}

\newcommand{\eps}{\varepsilon}
\renewcommand{\rho}{\varrho}

\newcommand{\normsq}[1]{\left\lVert #1 \right\rVert^2}
\newcommand{\norm}[1]{\left\lVert #1 \right\rVert}
\newcommand{\norminf}[1]{\left\lVert #1 \right\rVert_{\infty}}

\newcommand{\diag}[1]{\textnormal{diag}\left( #1 \right)}
\newcommand{\rank}[1]{\textnormal{rank}\left( #1 \right)}
\newcommand{\rankest}[1]{\textnormal{rank}_{\rho}\left( #1 \right)}
\renewcommand{\text}[1]{\textnormal{#1}}
\newcommand{\col}[1]{\textnormal{col}\left(#1\right)}
\newcommand{\lag}[1]{\textnormal{lag}\left(#1\right)}

\usepackage[acronym, style=alttree, toc=true, shortcuts, xindy, nomain, nonumberlist]{glossaries}

\newglossary{symbols}{sym}{sbl}{List of Symbols}
\newglossary{notation}{not}{nt}{Notation}

\glsaddkey{dimension}{\glsentrytext{\glslabel}}{\glsentryd}{\GLsentryd}{\glsd}{\Glsd}{\GLSd}

\newacronym{MPC}{MPC}{model predictive control}
\newacronym{iss}{ISS}{input-to-state stability}

\newglossaryentry{x}{type=symbols,
	sort={x},
	dimension={\ensuremath{ \mathbb{R}^{n} }},
	name={\ensuremath{\bm{x}}},
	description={State}
}

\newglossaryentry{xh}{type=symbols,
	sort={x},
	dimension={\ensuremath{ \mathbb{R}^{n} }},
	name={\ensuremath{\hat{\bm{x}}}},
	description={Measured state}
}

\newglossaryentry{u}{type=symbols,
	sort={u},
	dimension={\ensuremath{ \mathbb{R}^{m} }},
	name={\ensuremath{\bm{u}}},
	description={Input}
}

\newglossaryentry{y}{type=symbols,
	sort={y},
	dimension={\ensuremath{ \mathbb{R}^{r}}},
	name={\ensuremath{\bm{y}}},
	description={Output}
}

\newglossaryentry{d}{type=symbols,
	sort={d},
	dimension={\ensuremath{ \mathbb{D} }},
	name={\ensuremath{\bm{d}}},
	description={Disturbance}
}

\newglossaryentry{w}{type=symbols,
	sort={w},
	dimension={\ensuremath{ \mathbb{W} }},
	name={\ensuremath{\bm{w}}},
	description={General / extended disturbance}
}

\newglossaryentry{eps}{type=symbols,
	sort={e},
	dimension={\ensuremath{ \mathbb{E}}},
	name={\ensuremath{\bm{\eps}}},
	description={Measurement noise}
}

\newglossaryentry{ud}{type=symbols,
	sort={u},
	dimension={\ensuremath{ \mathcal{U}^{\textnormal{d}} }},
	name={\ensuremath{\bm{u}^{\textnormal{d}}}},
	description={Input data}
}

\newglossaryentry{yd}{type=symbols,
	sort={y},
	dimension={\ensuremath{ \mathcal{Y}^{\textnormal{d}} }},
	name={\ensuremath{\bm{y}^{\textnormal{d}}}},
	description={Output data}
}

\newglossaryentry{udt}{type=symbols,
	sort={u},
	dimension={\ensuremath{ \widetilde{\mathcal{U}}^{\textnormal{d}} }},
	name={\ensuremath{\widetilde{\bm{u}}^{\textnormal{d}}}},
	description={Input data}
}

\newglossaryentry{ydt}{type=symbols,
	sort={y},
	dimension={\ensuremath{ \widetilde{\mathcal{Y}}^{\textnormal{d}} }},
	name={\ensuremath{\widetilde{\bm{y}}^{\textnormal{d}}}},
	description={Output data}
}

\newglossaryentry{up}{type=symbols,
	sort={u},
	dimension={\ensuremath{ \mathcal{U}^{\textnormal{d}} }},
	name={\ensuremath{\widetilde{\bm{u}}}},
	description={Past input trajectory}
}

\newglossaryentry{yp}{type=symbols,
	sort={y},
	dimension={\ensuremath{ \mathcal{Y}^{\textnormal{d}} }},
	name={\ensuremath{\widetilde{\bm{y}}}},
	description={Past output trajectory}
}

\newglossaryentry{epsd}{type=symbols,
	sort={\eps},
	dimension={\ensuremath{ \mathbb{R}^{n} }},
	name={\ensuremath{\bm{\eps}^{\textnormal{d}}}},
	description={Measurement noise data}
}

\begin{document}
\begin{frontmatter}

\title{An Online Adaptation Strategy for \\ Direct Data-driven Control\thanksref{footnoteinfo}} 

\thanks[footnoteinfo]{\copyright~2023 the authors. This work has been accepted to IFAC for publication under a Creative Commons Licence CC-BY-NC-ND}

\author[First]{Johannes Teutsch} 
\author[First]{Sebastian Ellmaier} 
\author[First]{Sebastian Kerz}
\author[First]{Dirk Wollherr}
\author[First]{Marion Leibold}

\address[First]{Technical University of Munich, Department of Computer Engineering, Chair of Automatic Control Engineering (LSR), Theresienstra{\ss}e 90, 80333 Munich, Germany (e-mail: { \{johannes.teutsch, sebastian.ellmaier, s.kerz, dw, marion.leibold\}@tum.de})}

\begin{abstract}                
The fundamental lemma from behavioral systems theory yields a data-driven non-parametric system representation that has shown great potential for the data-efficient control of unknown linear and weakly nonlinear systems, even in the presence of measurement noise. In this work, we strive to extend the applicability of this paradigm to more strongly nonlinear systems by updating the system representation during control. Unlike existing approaches, our method does not impose suitable excitation to the control inputs, but runs as an observer parallel to the controller. Whenever a rank condition is deemed to be fulfilled, the system representation is updated using newly available datapoints. In a reference tracking simulation of a two-link robotic arm, we showcase the performance of the proposed strategy in a predictive control framework. 
\end{abstract}


\end{frontmatter}

\section{INTRODUCTION}
\vspace{-2mm}
The fundamental lemma from \cite{willems2005note} is a powerful result from behavioral systems theory that provides an appealing alternative to classical system models. The lemma allows to characterize the space of all fixed-length input-output trajectories of a linear time-invariant (LTI) system as the column span of Hankel matrices. These Hankel matrices consist of an input-output trajectory that was generated by a persistently exciting input signal (i.e., the input data matrix is of sufficient rank). The resulting non-parametric system representation has inspired many approaches for solving analysis and control problems directly from data, without the necessity of model identification~\citep{markovsky2008,datainformativity,berberichECC2020}; see the broad recent review by~\cite{behavioraltheory2021}.

When it comes to controller design, Willems' fundamental lemma allows for deriving stabilizing controllers directly from data, e.g., stabilizing state-feedback controllers and Linear Quadratic Regulators (LQR) \citep{de2019formulas}, as well as predictive controllers \citep{ACC2015,coulson2019data,berberich2020data}. Particularly in predictive control schemes, the data-driven system representation is used to predict and optimize over future input-output sequences, thus substituting the classical prediction model in Model Predictive Control (MPC). Such data-driven predictive control schemes are commonly known as Data-enabled Predictive Control (DeePC)~\citep{coulson2019data} or Data-driven Model Predictive Control (DD-MPC)~\citep{berberich2020data}. 

A major drawback of the fundamental lemma is that it holds only for deterministic LTI systems with exact data. If the data-generating system yields corrupted data due to noise or nonlinearity, the fundamental lemma is no longer valid. Nevertheless, DD-MPC schemes based on slightly modified formulations, using slack variables and regularization techniques, showed remarkable performance when applied to noisy and weakly nonlinear systems \citep{deepcQuadcop,berberichAT}, as well as time-varying systems \citep{baros2022online}. For LTI systems subject to bounded measurement noise, \cite{berberich2020data} provided the first theoretical analysis of stability and robustness of a DD-MPC scheme. For special classes of nonlinear systems, extensions of the fundamental lemma have been developed which can be used instead, e.g., for flat nonlinear systems \citep{alsalti2021data} and polynomial systems \citep{markovsky2021data}. A kernel-based extension to general nonlinear systems was recently provided by \cite{huang2022robust}, however, the prediction is only accurate locally where offline data have been collected.

\cite{berberich2022linear} showed that DD-MPC can be applied to weakly nonlinear systems by continuously updating the Hankel matrices with online data, with the idea that the data-driven system representation acts as a local linearization of the nonlinear system. In every time-step, the oldest datapoint in the Hankel matrix is discarded, all data is shifted back by one step, and the new most recent datapoint is attached at the end. However, continuously updating the data matrix is only valid if the inputs are persistently exciting, which cannot be guaranteed in closed-loop, especially when tracking a setpoint. To ensure persistently exciting data, \cite{berberich2022linear} propose to add suitable excitation to the controller input during closed-loop operation, e.g., by injecting noise, which deteriorates control performance and puts unnecessary stress on actuators. Alternatively, data updates could be stopped once a neighborhood of a setpoint is reached, but restarting the data updates in case of system or setpoint changes is not directly possible with the Hankel structure, as a continuous input-output trajectory is required.

\textit{Contribution:}
In this work, we present an online adaptation strategy for the data-driven system representation in DD-MPC schemes that does not require imposing suitably exciting inputs to the system during control operation. Our approach allows for stopping data updates whenever a setpoint is reached, and restarting updates once the system is moving again. To stop and restart data updates, we exploit mosaic Hankel matrices~\citep{van2020willems} and a corresponding extension of the fundamental lemma which allows for using discontinuous input-output trajectories for the data-driven system representation. Our algorithm requires only a single initially measured persistently exciting input-output trajectory and updates the data-driven system representation in each time-step depending on an extended rank condition on the collected data matrix. To evaluate the rank condition in case of data that is corrupted by noise and/or nonlinearity, we make use of a Singular Value Decomposition (SVD) of the data matrices. Although we show the efficacy of our method in a DD-MPC framework, our approach does not solely apply to such predictive controllers, but can be used independently, for example in the data-driven design of state-feedback controllers~\citep{de2019formulas} where the feedback gain is recomputed in every control iteration.

\textit{Structure:}
The remainder of this work is structured as follows. Section~\ref{sec:prelim} gives an introduction to the data-driven framework and related results. In Section~\ref{sec:method}, we present the proposed online adaptation algorithm, and then evaluate the approach in a simulation study in Section~\ref{sec:eval}. The paper is concluded in Section~\ref{sec:conclusion}.

\textit{Notation:}
We write $\bm{0}$ for any zero matrix or vector, and sequences of vectors are shortened to $\bm{s}_{[a,\,b]} := \left\{\bm{s}_a,\,\dots,\,\bm{s}_b\right\}$. For a given ouput $\bm{y}_k$ at time-step $k$, we write $\bm{y}_{i|k}$ for the predicted output $i$ steps ahead. For any sequence of vectors $\bm{s}_{[1,\,T]}$ with length $T$, we define the corresponding Hankel matrix $\bm{H}_{L}\left(\bm{s}_{[1,\,T]}\right)$ of depth $L \le T$ as
\begin{equation} \label{eq:def_hankel}
\bm{H}_{L}\left(\bm{s}_{[1,\,T]}\right) = \mat{\bm{s}_1 & \bm{s}_2 & \cdots & \bm{s}_{T-L+1} \\ 
	\bm{s}_2 & \bm{s}_3 & \cdots & \bm{s}_{T-L+2} \\
	\vdots & \vdots & \ddots & \vdots \\
	\bm{s}_{L} & \bm{s}_{L+1} & \cdots & \bm{s}_{T}}.
\end{equation}
By $\col{\bm{s}_a,\,\ldots,\,\bm{s}_b} := \vc{\bm{s}^{\top}_a,\,\ldots,\,\bm{s}^{\top}_b}^{\top}$, we denote the result from stacking the vectors/matrices $\bm{s}_a,\,\ldots,\,\bm{s}_b$. For a vector $\bm{s}$ and a matrix $\bm{P}$, we define the weighted 2-norm of $\bm{s}$ as $\norm{\bm{s}}_{\bm{P}} := \sqrt{\bm{s}^{\top} \bm{P} \bm{s}}$.

\section{DATA-DRIVEN SYSTEM REPRESENTATIONS} \label{sec:prelim}
In this section, we present results of behavioral systems theory that provide data-driven system representations for discrete-time LTI systems.

Consider a discrete-time LTI system $G$ with unknown state-space parameters $\bm{A}$, $\bm{B}$, $\bm{C}$, $\bm{D}$, i.e.,
\begin{subequations}\label{eq:ltisystem}
	\begin{align}
	\gls{x}_{k+1} &= \bm{A} \gls{x}_{k} + \bm{B} \gls{u}_{k}, \label{eq:system1}\\
	\gls{y}_{k} &= \bm{C} \gls{x}_{k} + \bm{D} \gls{u}_{k}
	\end{align}
\end{subequations}
where the state, input, and output are denoted as $\gls{x}_{k} \in \glsd{x}$, $\gls{u}_{k} \in \glsd{u}$, and $\gls{y}_{k} \in \glsd{y}$, respectively. The following result from behavioral systems theory, known as the \textit{fundamental lemma}, allows for a non-parametric representation of \eqref{eq:ltisystem} based on input-output data.
\begin{lem}[\cite{willems2005note}]\label{lem:fundamental}
	Consider a controllable LTI system $G$ of the form \eqref{eq:ltisystem} and a given data trajectory $(\gls{ud}_{[1,\,T]},~\gls{yd}_{[1,\,T]})$ of length $T \ge L := T_{\text{p}} + T_{\text{f}}$, with time windows $T_{\text{p}} \ge \lag{G}$\footnote{$\lag{G}$ of an LTI system $G$ is the smallest natural number $j \le n$ for which the observability matrix $\col{\bm{C}, \bm{C} \bm{A}, \ldots, \bm{C} \bm{A}^{j-1}}$ has rank $n$.}, $T_{\text{f}} \ge 1$. If
	\begin{equation} \label{eq:perexc}
	    \rank{\bm{H}_{n+L}\left(\gls{ud}_{[1,\,T]}\right)} = m \left(n+L\right),
	\end{equation}
	then any $L$-long input-output sequence $\left(\gls{u}_{[k-T_{\text{p}},\,k+T_{\text{f}}-1]},\right.$ $\left.\gls{y}_{[k-T_{\text{p}},\,k+T_{\text{f}}-1]}\right)$ is a valid trajectory of $G$ for all $k \ge T_{\text{p}}$ if and only if there exists an $\bm{\alpha} \in \mathbb{R}^{T-L+1}$ such that
	\begin{equation} \label{eq:fundlemm}
		\vc{\col{\gls{u}_{k-T_{\text{p}}},\,\ldots,\,\gls{u}_{k+T_{\text{f}}-1}}\\ \col{\gls{y}_{k-T_{\text{p}}},\,\ldots,\,\gls{y}_{k+T_{\text{f}}-1}}} = \mat{\bm{H}_{L}\left(\gls{ud}_{[1,\,T]}\right) \\ \bm{H}_{L}\left(\gls{yd}_{[1,\,T]}\right)} \bm{\alpha}.
	\end{equation}
\end{lem}
If the input data $\gls{ud}_{[1,\,T]}$ satisfies \eqref{eq:perexc}, $\gls{ud}_{[1,\,T]}$ is said to be \textit{persistently exciting of order} $n+L$.

With \eqref{eq:fundlemm}, Lemma~\ref{lem:fundamental} provides a data-driven description of all possible input-output trajectories of a controllable LTI system by only using a previously recorded trajectory of finite length $T$ with persistently exciting input data $\gls{ud}_{[1,\,T]}$. Thus, \eqref{eq:fundlemm} can be used as a non-parametric predictive representation for system \eqref{eq:ltisystem}, e.g., in predictive control schemes~\citep{coulson2019data, berberich2020data}. Note that the past input-output trajectory $\left(\gls{u}_{[k-T_{\text{p}},\,k-1]},\,\gls{y}_{[k-T_{\text{p}},\,k-1]}\right)$ is used in \eqref{eq:fundlemm} to fix the initial state of the system in order to retrieve uniquely determined predictions~\citep{markovsky2008}, and that $T \ge (m+1)(n+T_{\text{p}}+T_{\text{f}}) - 1$ is necessary for \eqref{eq:perexc}.

Lemma~\ref{lem:fundamental} relies on a persistency of excitation condition on the input data with an order that is larger than the length of the predicted trajectory. In \cite{willems2005note}, this condition was originally shown to be only sufficient for \eqref{eq:fundlemm}, however, \cite{berberich2022quantitative} have recently shown that this condition is also necessary when considering controllable LTI systems. 

Lemma~\ref{lem:fundamental} can be extended to more general data structures than Hankel matrices, allowing for a dataset consisting of multiple short data trajectories instead of one long trajectory. Before we present the extension of Lemma~\ref{lem:fundamental}, we first define the generalized \textit{mosaic} Hankel matrix.
\begin{defn}[\cite{van2020willems}] \label{def:mosaicHankel}
~~~~~Let $\mathcal{S}=$ $ \{\bm{s}^{1}_{[1,T_1]},\,\ldots,\,\bm{s}^{N}_{[1,T_N]}\}$ be the set of $N$ vector sequences with lengths $T_1,\,\ldots,\,T_N$. We define the corresponding mosaic Hankel matrix $\bm{\mathcal{H}}_{L}\left(\mathcal{S}\right)$ of depth $L \le \min(T_1,\,\ldots,\,T_N)$ as
\begin{equation} \label{eq:def_mosaicHankel}
\bm{\mathcal{H}}_{L}\left(\mathcal{S}\right) = \mat{\bm{H}_{L}\left(\bm{s}^{1}_{[1,\,T_1]}\right),\,\ldots,\,\bm{H}_{L}\left(\bm{s}^{N}_{[1,\,T_N]}\right)}.
\end{equation}
\end{defn}
Note that the Hankel matrix \eqref{eq:def_hankel} is a special case of the mosaic Hankel matrix \eqref{eq:def_mosaicHankel}. The general form \eqref{eq:def_mosaicHankel} also includes other special forms like Page or trajectory matrices~\citep{behavioraltheory2021}.

The following lemma extends Lemma~\ref{lem:fundamental} to mosaic Hankel matrices.
\begin{lem}[\cite{markovsky2022identifiability}]\label{lem:extfundamental}
	Consider an \linebreak LTI system $G$ of the form \eqref{eq:ltisystem} and given datasets $\glsd{ud},\, \glsd{yd}$ consisting of $N \ge 1$ input-output trajectories of lengths $T_1,\ldots,T_N \ge L := T_{\text{p}} + T_{\text{f}}$, with time windows \linebreak $T_{\text{p}} \ge \lag{G}$, $T_{\text{f}} \ge 1$. If
	\begin{equation} \label{eq:perexc_ext}
	    \rank{\mat{\bm{\mathcal{H}}_{L}\left(\glsd{ud}\right)\\ \bm{\mathcal{H}}_{L}\left(\glsd{yd}\right)}} = n + m L,
	\end{equation}
	then any $L$-long input-output sequence $\left(\gls{u}_{[k-T_{\text{p}},\,k+T_{\text{f}}-1]},\right.$ $\left.\gls{y}_{[k-T_{\text{p}},\,k+T_{\text{f}}-1]}\right)$ is a valid trajectory of $G$ for all $k \ge T_{\text{p}}$ if and only if there exists an $\bm{\alpha} \in \mathbb{R}^{\Sigma_{i=1}^N (T_i-L+1)}$ such that
	\begin{equation} \label{eq:extfundlemm}
		\vc{\col{\gls{u}_{k-T_{\text{p}}},\,\ldots,\,\gls{u}_{k+T_{\text{f}}-1}}\\ \col{\gls{y}_{k-T_{\text{p}}},\,\ldots,\,\gls{y}_{k+T_{\text{f}}-1}}} = \mat{\bm{\mathcal{H}}_{L}\left(\glsd{ud}\right) \\ \bm{\mathcal{H}}_{L}\left(\glsd{yd}\right)} \bm{\alpha}.
	\end{equation}
\end{lem}

The main differences between Lemma~\ref{lem:fundamental} and Lemma~\ref{lem:extfundamental} are 
a) Lemma~\ref{lem:extfundamental} considers the general data matrix structure \eqref{eq:def_mosaicHankel}, which allows for using multiple short (but sufficiently long) data trajectories instead of one long continuous trajectory,
b) Lemma~\ref{lem:extfundamental} considers both controllable and uncontrollable LTI systems, and
c) in Lemma~\ref{lem:extfundamental}, the excitation condition \eqref{eq:perexc} on the input data is replaced with the rank condition \eqref{eq:perexc_ext} on the input-output data matrix, also referred to as the \textit{generalized persistency of excitation} condition~\citep{behavioraltheory2021}.

Note that $\sum_{i=1}^{N} T_i \ge n + m(T_{\text{p}}+T_{\text{f}})+N(T_{\text{p}}+T_{\text{f}}-1)$ is necessary for \eqref{eq:perexc_ext}. For $T_i = T_{\text{p}}+T_{\text{f}}$, the lower bound on the number of data trajectories results in $N \ge n + m(T_{\text{p}}+T_{\text{f}})$. A drawback of Lemma~\ref{lem:extfundamental} is that the system order $n$ must be exactly known due to condition \eqref{eq:perexc_ext}. In comparison, Lemma~\ref{lem:fundamental} is also valid when using an upper bound $\ol{n} \ge n$ for \eqref{eq:perexc}, e.g., $\ol{n} = p\, \lag{G}$~\citep{markovsky2022identifiability}.

Both Lemma~\ref{lem:fundamental} and Lemma~\ref{lem:extfundamental} hold for deterministic LTI systems \eqref{eq:ltisystem}. If the data-generating system yields corrupted data due to noise or nonlinearity, \eqref{eq:fundlemm} and \eqref{eq:extfundlemm} do not hold. Despite this fact, \cite{berberich2022linear} showed that DD-MPC schemes based on Lemma~\ref{lem:fundamental} can be successfully applied to weakly nonlinear systems when the used dataset is continuously updated with a fixed size moving window, such that \eqref{eq:fundlemm} acts as a local linearization of the system. However, continuously updating the data matrix in \eqref{eq:fundlemm} is only valid if the inputs satisfy \eqref{eq:perexc}, and applying suitable excitation inputs in closed-loop is not always desired during the control phase, e.g., when setpoints are reached. In the following section, we address this problem by deriving an online adaptation strategy based on Lemma~\ref{lem:extfundamental}.

\section{Online Adaptation Strategy} \label{sec:method}
The key property of Lemma~\ref{lem:extfundamental} that we will make use of is that a mosaic Hankel matrix \eqref{eq:def_mosaicHankel} of the input-output data can be used. 
This allows for integration of not only one long continuous input-output trajectory as in \eqref{eq:def_hankel}, but also of discontinuous trajectories as far as each reduced continuous stretch is sufficiently long.
In fact, we can interpret each column in \eqref{eq:def_mosaicHankel} as a separate $L$-long trajectory of the system. This interpretation is the foundation of our online adaptation strategy; append the latest $L$-long trajectory to the dataset and discard the oldest trajectory, only if the rank condition \eqref{eq:perexc_ext} is still fulfilled.
\subsection{Coping with Corrupted Data}
\begin{figure}
    \centering
    \def\svgwidth{0.49\textwidth}
    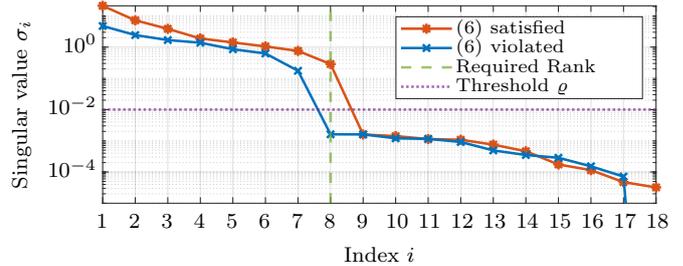
    \vspace*{-7mm}
    \caption{Exemplary comparison of singular values from noisy data matrices whose uncorrupted counterparts satisfy or violate rank condition \eqref{eq:perexc_ext}. Thresholding the singular values with $\rho$ recovers the original rank.}
    \label{fig:svd}
\end{figure}

For deterministic LTI systems \eqref{eq:ltisystem}, the equality in \eqref{eq:perexc_ext} only holds if the data are sufficiently informative. If that is not the case, then the rank of the data matrix is lower than required in \eqref{eq:perexc_ext}. However, in case of corrupted data, the rank might even increase as the data cannot be explained by an LTI system $G$ of complexity $\left(m,\lag{G},n\right)$ \citep{behavioraltheory2021}. To overcome this issue, \cite{coulson2022robust} have proposed a quantitative measure of persistency of excitation by lower-bounding the minimum singular value of the input data matrix, thus denoting input-output data matrices that are more robust to noise. Motivated by \cite{coulson2022robust}, we make use of SVD to define an alternative rank function in order to cope with corrupted data, instead \linebreak of evaluating the fragile rank condition \eqref{eq:perexc_ext} directly.
\begin{defn}[Robustified rank]\label{def:robrank}
    ~~Consider a matrix \linebreak $\bm{M} \in \mathbb{R}^{v \times q}$ and a threshold $\rho \ge 0$. We define the robustified rank $\rankest{M}$ as the number of singular values $\sigma_1,\,\ldots,\,\sigma_{\min(v,q)}$ that are greater than $\rho$.
\end{defn}
Fig.~\ref{fig:svd} shows singular values of an exemplary noisy data matrix $\col{\bm{\mathcal{H}}_{T_{\text{p}}+T_{\text{f}}}\left(\glsd{ud}\right),\, \bm{\mathcal{H}}_{T_{\text{p}}+T_{\text{f}}}\left(\glsd{yd}\right)}$ for the cases where the noise-free counterparts satisfy or violate the rank condition \eqref{eq:perexc_ext}. Due to the noise, the rank of the data matrix exceeds the bound in \eqref{eq:perexc_ext} for both shown cases. However, when using the robustified rank function from Def.~\ref{def:robrank}, the uncorrupted rank of the noise-free data matrix can be obtained by a suitable choice of the threshold $\rho$. Note that this is only possible if the signal-to-noise ratio (SNR) is adequate, i.e., there exists a threshold $\rho$ that is able to correctly divide singular values stemming from informative data and the corrupting noise. For $\rho = 0$, it immediately follows that $\rankest{\cdot} \equiv \text{rank}(\cdot)$.

In the literature on truncated SVD, there exist optimal truncation thresholds for matrices $\bm{M}$ whose elements are affected by independent and identically distributed Gaussian noise with known magnitude \citep{brunton2022data}. However, these thresholds are not applicable in our case, as there generally exist multiple identical matrix elements in the used mosaic Hankel matrix \eqref{eq:def_mosaicHankel}. Determining a suitable threshold $\rho$ for general data matrices as in \eqref{eq:extfundlemm} is an interesting open problem for research. Thus, in the following, we assume that a suitable threshold is known apriori, e.g., through experiments and data analysis. A possible heuristic to determine such thresholds is to first apply exciting (e.g., random) inputs to the system to retrieve sufficiently informative input-output data, and then plot the singular values of the data matrix, akin to Fig.~\ref{fig:svd}: given the required rank~\eqref{eq:perexc_ext}, the threshold $\rho$ must be chosen such that $\sigma_{n + m \left(T_{\text{p}}+T_{\text{f}}\right)} \ge \rho \ge \sigma_{n + m \left(T_{\text{p}}+T_{\text{f}}\right)+1}$ holds.

\subsection{Online Adaptation Algorithm}
We now describe the proposed strategy for online adaptation of the data-driven system representation \eqref{eq:extfundlemm}.

Let $\glsd{ud}_0 =\{ \bm{u}^{\text{d},1}_{[1,L]},\ldots,\bm{u}^{\text{d},N}_{[1,L]}\}$, $\glsd{yd}_0 =\{ \bm{y}^{\text{d},1}_{[1,L]},\ldots,\bm{y}^{\text{d},N}_{[1,L]}\}$ be an initial dataset of $N \ge n + m L$ sufficiently informative input-output trajectories of length $L := T_{\text{p}} + T_{\text{f}}$ with sufficiently large SNR, and let $\rho > 0$ be an appropriately chosen threshold. Furthermore, we define the most recent $L$-long input-output trajectory at time-step $k$ as
\begin{subequations} \label{eq:recent-io}
\begin{align}
    \gls{up}_k := \col{\gls{u}_{k-T_\text{p}-T_\text{f}},\,\ldots,\,\gls{u}_{k-1}},\\
    \gls{yp}_k := \col{\gls{y}_{k-T_\text{p}-T_\text{f}},\,\ldots,\,\gls{y}_{k-1}}.
\end{align}
\end{subequations}

Our algorithm works as follows. First, the past trajectory $\gls{up}_0,\,\gls{yp}_0$ needs to be initialized. If the control phase starts immediately after the data collection phase for the initial datasets $\glsd{ud}_0$, $\glsd{yd}_0$, the initial past trajectory $\gls{up}_0,\,\gls{yp}_0$ is set to the most recent data. If there is no past input-output trajectory of length $L$ available at $k=0$, an artificial steady-state trajectory can be used, best corresponding to the initial state of the system. If the values for input and output at the initial state are unknown, any constant values can be chosen, with the only drawback of an $L$-step warm-up phase in which predictions do not match reality.
This discrepancy vanishes as soon as $L$ input-output datapoints are measured. 

Once the past trajectory $\gls{up}_0,\,\gls{yp}_0$ is initialized, the control loop begins: In each time-step $k\ge0$, the control input $\gls{u}_k$ is applied (determined by, e.g., a DD-MPC scheme using the current dataset $\glsd{ud}_k$, $\glsd{yd}_k$, as in \cite{berberich2022linear}). After the resulting output $\gls{y}_k$ is measured, we construct candidate datasets $\glsd{udt}_{k+1}$, $\glsd{ydt}_{k+1}$ for the following time-step by discarding the oldest data trajectory $\bm{u}^{\text{d},1}_{[1,L]}$, $\bm{y}^{\text{d},1}_{[1,L]}$ and appending the most recent trajectory \eqref{eq:recent-io}, i.e.,
\begin{subequations} \label{eq:data_cand}
\begin{align}
    \glsd{udt}_{k+1} &=\left\{ \bm{u}^{\text{d},2}_{[1,L]},\,\ldots,\,\bm{u}^{\text{d},N}_{[1,L]},\, \gls{up}_{k+1}\right\},\\
    \glsd{ydt}_{k+1} &= \left\{ \bm{y}^{\text{d},2}_{[1,L]},\,\ldots,\,\bm{y}^{\text{d},N}_{[1,L]},\,\gls{yp}_{k+1}\right\}.
\end{align}
\end{subequations}
Then, we construct a mosaic Hankel matrix \eqref{eq:def_mosaicHankel} using the candidate datasets \eqref{eq:data_cand} and use $\rankest{\cdot}$ to evaluate the rank condition \eqref{eq:perexc_ext}, yielding the following two possibilities:
\begin{subequations}\label{eq:cases}
\begin{align}
 \rankest{\col{\bm{\mathcal{H}}_{L}\left(\glsd{udt}_{k+1}\right),\, \bm{\mathcal{H}}_{L}\left(\glsd{ydt}_{k+1}\right)}} &< n + mL, \label{eq:caselow}\\
 \rankest{\col{\bm{\mathcal{H}}_{L}\left(\glsd{udt}_{k+1}\right),\, \bm{\mathcal{H}}_{L}\left(\glsd{ydt}_{k+1}\right)}} &\ge n + mL. \label{eq:casehigh}
\end{align}
\end{subequations}
If \eqref{eq:caselow} holds, then the data \eqref{eq:data_cand} are believed not to be sufficiently informative, and thus are invalid for the data-driven system representation \eqref{eq:extfundlemm}. Therefore, we choose the actually used dataset for the next time-step $k+1$ to be $\glsd{ud}_{k+1} := \glsd{ud}_k$, $\glsd{yd}_{k+1} := \glsd{yd}_k$. In contrast, if \eqref{eq:casehigh} holds, then the data \eqref{eq:data_cand} are believed to be sufficiently informative, and thus can be used for the data-driven system representation \eqref{eq:extfundlemm}. Therefore, we choose the actually used dataset for the next time-step $k+1$ to be $\glsd{ud}_{k+1} := \glsd{udt}_{k+1}$, $\glsd{yd}_{k+1} := \glsd{ydt}_{k+1}$. We summarize the proposed strategy in Algorithm~\ref{alg:method}.
\begin{algorithm}
\caption{Online Adaptation}\label{alg:method}
\begin{algorithmic}[1]
\Require Initial datasets $\glsd{ud}_0$, $\glsd{yd}_0$, system order $n$, input dimension $m$, time windows $T_{\text{p}}$, $T_{\text{f}}$, threshold $\rho$.

\State Initialize $\gls{up}_0$, $\gls{yp}_0$
\For{$k \ge 0$}
\State Apply controller using current dataset $\glsd{ud}_k$, $\glsd{yd}_k$
\State Update recent trajectory $\gls{up}_{k+1}$, $\gls{yp}_{k+1}$ of length $L$
\State Construct candidate dataset $\glsd{udt}_{k+1}$, $\glsd{ydt}_{k+1}$ via \eqref{eq:data_cand}
\State Evaluate rank condition \eqref{eq:cases}:
\If{\eqref{eq:caselow}}
    \State Set $\glsd{ud}_{k+1} := \glsd{ud}_k$, $\glsd{yd}_{k+1} := \glsd{yd}_k$
\ElsIf{\eqref{eq:casehigh}}
    \State Set $\glsd{ud}_{k+1} := \glsd{udt}_{k+1}$, $\glsd{yd}_{k+1} := \glsd{ydt}_{k+1}$
\EndIf
\EndFor
\end{algorithmic}
\end{algorithm}

For deterministic LTI systems and $\rho = 0$, Algorithm~\ref{alg:method} can be combined with Lemma~\ref{lem:extfundamental} without loosing validity, as the rank condition \eqref{eq:perexc_ext} holds for all $k \ge 0$.

\section{NUMERICAL EVALUATION} \label{sec:eval}
In this section, we evaluate the proposed approach in a reference tracking simulation of a two-link robotic arm.

\subsection{Simulation Setup}
\begin{figure}
    \centering
    \vspace*{2mm}
    \newcommand{\nvar}[2]{%
    \newlength{#1}
    \setlength{#1}{#2}
}

\nvar{\dg}{0.2cm}
\def\dw{0.25}
\nvar{\ddx}{1cm}

\def\link{
    \draw [distance=1.5mm, very thick] (0,0)--
}
\def\joint{%
    \filldraw [ultra thick, fill=white] (0,0) circle (3pt);
}
\def\grip{%
    \draw[ultra thick](0cm,\dg)--(0cm,-\dg);
    \draw[ultra thick] (0cm, \dg)--(\dg,\dg);
    \draw[ultra thick] (0cm, -\dg)--(\dg,-\dg);
}
\def\robotbase{%
    \draw[thick,-latex] (0,0) -- (0,1.7) node[anchor=east] {$y$};
    \draw[thick,-latex] (0,0) -- (1.7,0) node[anchor=north] {$x$};
  
}

\newcommand{\angann}[2]{%
    \begin{scope}
    \draw [dashed] (0,0) -- (1.2\ddx,0pt);
    \draw [->, shorten >=3.5pt] (\ddx,0pt) arc (0:#1:\ddx);
    \node at (#1/2-2:\ddx+8pt) {#2};
    
    \end{scope}
}

\newcommand{\trqann}[1]{%
    \begin{scope}
    \draw [thick, ->] (-0.866*3mm,0.5*3mm) arc (150:30:3mm);
    \node at (0,6mm) {#1};
    
    \end{scope}
}

\newcommand{\lineann}[5][0.5]{%
    \begin{scope}[rotate=#2, inner sep=2pt]
        \draw (#3/2, #1) node[fill=white] {#4};
        \draw (0.9*#3, 1.2*#1) node[fill=white] {#5};
    \end{scope}
}

\def\thetaone{-55}
\def\Lone{2.2}
\def\thetatwo{-50}
\def\Ltwo{1.4}
\small
\begin{tikzpicture}
    \robotbase
    \begin{scope}[rotate = 90]
    \angann{\thetaone}{$\theta_1$}
    \trqann{$\tau_1$}
    \lineann[-0.3]{\thetaone}{\Lone}{$l_1$}{$m_1$}
    \link(\thetaone:\Lone);
    \joint
    \begin{scope}[shift=(\thetaone:\Lone), rotate=\thetaone]
        \angann{\thetatwo}{$\theta_2$}
        \trqann{$\tau_2$}
        \lineann[-0.3]{\thetatwo}{\Ltwo}{$l_2$}{$m_2$}
        \link(\thetatwo:\Ltwo);
        \joint
        \begin{scope}[shift=(\thetatwo:\Ltwo), rotate=\thetatwo]
          \grip
        \end{scope}
    \end{scope}
    \end{scope}
    
    \draw[thick, -latex] (4.5,1.7) -- (4.5,1);
    \node at (4.7,1.4) {$g$};
    
\end{tikzpicture}
    \caption{Schematic illustration of the two-link robot.}
    \label{fig:robot}
\end{figure}
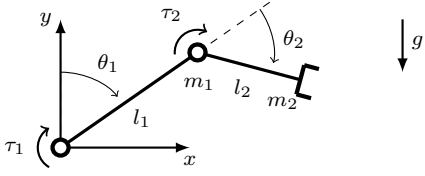

We consider a two-link robotic arm with the dynamics
\begin{equation}\label{eq:robotmodel}
    \bm{M}\left(\bm{\theta}\right) \ddot{\bm{\theta}} + \bm{C}\left(\bm{\theta},\dot{\bm{\theta}}\right) + \bm{G}\left(\bm{\theta}\right) = \bm{\tau},
\end{equation}
where $\bm{\theta} = \col{\theta_1,\, \theta_2}$ are the joint angles and $\bm{\tau} = \col{\tau_1,\, \tau_2}$ are the joint torques, and where
\begin{subequations}
\begin{align*}
    \bm{M}\left(\bm{\theta}\right) &= \mat{\tilde{m} l_1^2 + m_2 l_2^2 + 2 m_2 l_1 l_2 \cos(\theta_2) & * \\  m_2 l_2^2 + m_2 l_1 l_2 \cos(\theta_2) & m_2 l_2^2},\\
    \bm{C}\left(\bm{\theta},\dot{\bm{\theta}}\right) &= \vc{-m_2 l_1 l_2 \left(2 \dot{\theta}_1 \dot{\theta}_2 + \dot{\theta}_2^2\right) \sin(\theta_2) + d_1 \dot{\theta}_1\\ -m_2 l_1 l_2 \dot{\theta}_1 \dot{\theta}_2 \sin(\theta_2) + d_2 \dot{\theta}_2} , \\
    \bm{G}\left(\bm{\theta}\right) &= \vc{-\tilde{m} g l_1 \sin(\theta_1) - m_2 g l_2 \sin(\theta_1 + \theta_2) \\ -m_2 g l_2 \sin(\theta_1 + \theta_2)},
\end{align*}
\end{subequations}
are the inertia matrix, the vector of Coriolis and centrifugal forces, and the vector of gravity torques, respectively. Note that $*$ represents matrix entries that results from symmetry, and $\tilde{m} = m_1 + m_2$. System~\eqref{eq:robotmodel} is determined by the masses $m_1 = \SI{0.3}{\kilo\gram}$, $m_2 = \SI{0.1}{\kilo\gram}$, the arm lengths $l_1 = \SI{0.4}{\meter}$, $l_2 = \SI{0.2}{\meter}$, the damping constants $d_1 = d_2 = \SI{0.001}{\kilo\gram\,\meter ^2/\second}$, and the acceleration of gravity $g =$ $ \SI{9.81}{\kilo\gram\,\meter ^2/\second^2}$. Fig.~\ref{fig:robot} illustrates the robot configuration.

With a sampling time of $\D t = \SI{0.01}{\second}$, we define the discrete-time input and output of \eqref{eq:robotmodel} as $\gls{u}_k = \bm{\tau}_k$ and $\gls{y}_k = \bm{\theta}_k + \gls{eps}_k$, with the measurement noise $\gls{eps}_k \in \glsd{eps}$, $\glsd{eps} = \left\{ \gls{eps} \in \mathbb{R}^2 ~\left|~ \norminf{\gls{eps}} \le 10^{-3} \right.\right\}$. Note that the discrete-time input $\gls{u}_k$ is applied to \eqref{eq:robotmodel} in zero-order hold fashion.

The goal is to track a reference trajectory $\bm{y}^{\text{ref}}_{k}$ that starts at the lower equilibrium $\bm{\theta}^{\text{ini}} = \vc{-\pi,\,0}^{\top}$, steers to the intermediate joint position $\bm{\theta}^{\text{mid}} = \vc{-\pi/2,\,\pi/2}^{\top}$, and stays there for $\SI{3.5}{\second}$, before reaching the endpoint $\bm{\theta}^{\text{end}} = \bm{0}$.

\subsection{Controller Design}
In order to achieve the control goal without model knowledge, we combine the proposed approach for adapting the data-driven system representation with a DD-MPC scheme akin to~\cite{coulson2019data}. We define the optimal control problem (OCP) that is solved in each time-step as
\begin{subequations} \label{eq:ocp}
\vspace*{-3mm}
	\begin{align}
		 &~\hspace{-9mm} \underset{\bm{u}_{\text{f},k},\, \bm{y}_{\text{f},k},\, \bm{\alpha},\, \bm{\mu}}{\text{minimize}}~ \sum\limits_{i=0}^{T_{\text{f}}-1} l\left(\gls{u}_{i|k},\, \gls{y}_{i|k}\right)
		+ \lambda_{\alpha} \norm{\bm{\alpha}}^2_2 + \lambda_{\mu} \norm{\bm{\mu}}^2_2 \label{eq:ocp_cost}\\	
		\text{s.t. }		
		& \mat{\bm{u}_{\text{p},k}\\
		\bm{u}_{\text{f},k}\\
		\bm{y}_{\text{p},k} + \bm{\mu}\\
		\bm{y}_{\text{f},k}} = \mat{\bm{\mathcal{H}}_{T_{\text{p}}+T_{\text{f}}}\left(\glsd{ud}_k\right) \\ \bm{\mathcal{H}}_{T_{\text{p}}+T_{\text{f}}}\left(\glsd{yd}_k\right)} \bm{\alpha}, \label{eq:ocp_hankel}\\
		& \norminf{\bm{u}_{\text{f},k}}\le u_{\text{max}},\\
		& \norminf{\D{} \gls{u}_{i|k}} \le \D{} u_{\text{max}} ~~\forall i \in \{0,\,\ldots,\,T_{\text{f}}-1\},
	\end{align}
\end{subequations}
with $\D{} \gls{u}_{i|k} := \gls{u}_{i|k} - \gls{u}_{i-1|k}$, $\gls{u}_{-1|k} := \gls{u}_{k-1}$, and where
\begin{subequations}
\begin{align}
    \bm{u}_{\text{f},k} :=  \col{\gls{u}_{0|k},\,\ldots,\,\gls{u}_{T_\text{f}-1|k}},\\
    \bm{y}_{\text{f},k} := \col{\gls{y}_{0|k},\,\ldots,\,\gls{y}_{T_\text{f}-1|k}},
\end{align}
\end{subequations}
denote the vectors of predicted $T_\text{f}$ in- and outputs, and
\begin{subequations}
\begin{align}
    \bm{u}_{\text{p},k} := \col{\gls{u}_{k-T_\text{p}},\,\ldots,\,\gls{u}_{k-1}},\\
    \bm{y}_{\text{p},k} := \col{\gls{y}_{k-T_\text{p}},\,\ldots,\,\gls{y}_{k-1}},
\end{align}
\end{subequations}
denote the vectors of past $T_\text{p}$ in- and outputs, respectively.
Note that the slack variable $\bm{\mu}$ is necessary to ensure feasibility of the constraint \eqref{eq:ocp_hankel}, as the past output-trajectory is corrupted by noise and nonlinearity, and thus might not lie in the image space of the data matrix $\col{\bm{\mathcal{H}}_{T_{\text{p}}+T_{\text{f}}}\left(\glsd{ud}_k\right),\, \bm{\mathcal{H}}_{T_{\text{p}}+T_{\text{f}}}\left(\glsd{yd}_k\right)}$~\citep{coulson2019data}. The truncated SVD of $\col{\bm{\mathcal{H}}_{T_{\text{p}}+T_{\text{f}}}\left(\glsd{ud}_k\right),\, \bm{\mathcal{H}}_{T_{\text{p}}+T_{\text{f}}}\left(\glsd{yd}_k\right)}$ can also be used in \eqref{eq:ocp_hankel}, which showed to potentially improve performance~\citep{coulson2019data}.

The stage cost $l\left(\gls{u}_{i|k}, \gls{y}_{i|k}\right)$ is defined as
\begin{equation}
    l\left(\gls{u}_{i|k},\gls{y}_{i|k}\right) =  \normsq{\gls{y}_{i|k}-\bm{y}^{\text{ref}}_{k+i}}_{\bm{Q}} + \normsq{\gls{u}_{i|k}}_{\bm{R}} + \normsq{\D{} \gls{u}_{i|k}}_{\bm{R}_{\D{}}},
\end{equation}
with the weights 
$\bm{Q} = \diag{1,1}$, $\bm{R} = 10^{-5} \diag{1,2}$, and $\bm{R}_{\D{}} = 10^{-4} \diag{2,4}$ 
for the predicted outputs, inputs, and rate of change of inputs, respectively. The input weights are chosen relatively small compared to the output weights such that larger, more likely exciting, inputs are not heavily penalized by the cost function. The predicted inputs $\bm{u}_{\text{f},k}$ and their rate of change $\D{} \gls{u}_{i|k}$ are bounded by $u_{\text{max}} = \SI{5}{\newton\meter}$ and $\D{} u_{\text{max}} = \SI{1}{\newton\meter}$. The regularization parameters in \eqref{eq:ocp_cost} are set to $\lambda_{\alpha} = 5\cdot 10^{-5}$ and $\lambda_{\mu} = 10^{3}$, and we choose the time windows to $T_\text{p} = 4$ and $T_\text{f} = 10$.

In each time-step $k$, the OCP \eqref{eq:ocp} is solved and the first input $\gls{u}^{\ast}_{0|k}$ from the optimal input vector $\bm{u}^{\ast}_{\text{f},k}$ is applied to system~\eqref{eq:robotmodel}, as common in predictive control schemes.

\subsection{Simulation Results}
We implement three different predictive control methods based on the OCP \eqref{eq:ocp} to evaluate and compare our proposed online adaptation strategy.
\begin{enumerate}
	\item \textit{Proposed Method} (PM): The proposed method updates the data $\glsd{ud}_k,\,\glsd{yd}_k$ in \eqref{eq:ocp_hankel} according to Algorithm~\ref{alg:method}.
	\item \textit{Always Update} (AU): This method updates the data $\glsd{ud}_k,\glsd{yd}_k$ in \eqref{eq:ocp_hankel} in each time-step. To ensure persistency of excitation, we add a random value $\gls{u}_k^{\text{rnd}}$ to the optimal input $\gls{u}^{\ast}_{0|k}$, where $\norminf{\gls{u}_k^{\text{rnd}}} \le \SI{0.25}{\newton\meter}$.
	\item \textit{Never Update} (NU): This method uses the initial dataset $\glsd{ud}_0,\glsd{yd}_0$ throughout the whole control phase.
\end{enumerate}
First, we collect the initial dataset $\glsd{ud}_0,\,\glsd{yd}_0$ by applying random inputs $\gls{ud}_k$, $\norminf{\gls{ud}_k} \le \SI{0.25}{\newton\meter}$, to system~\eqref{eq:robotmodel} in the lower equilibrium $\bm{\theta}^{\text{ini}}$ for $T = 55$ time-steps. This yields $N=42$ trajectories of length $L=14$ in Hankel structure \eqref{eq:def_hankel}. By inspection of the singular values of the initial data matrix, we choose the threshold for the robustified rank computation in Algorithm~\ref{alg:method} as $\rho = 0.005$.

The reference tracking simulation is then repeated $100$ times for each method with varying measurement noise $\gls{eps}_k$. We simulate $\SI{10}{\second}$ per run, corresponding to $1000$ time-steps. The simulations were carried out in MATLAB using the \texttt{quadprog} solver on an AMD Ryzen 5 Pro 3500U.

\begin{figure}
    \centering
    \def\svgwidth{0.415\textwidth}
    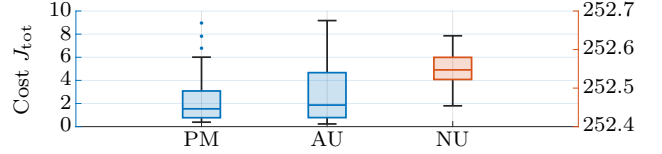
    \vspace*{-4mm}
    \caption{Boxplots of the total trajectory costs $J_{\text{tot}}$ for Proposed Method, Always Update, and Never Update.}
    \label{fig:cost}
\end{figure}

\begin{figure}
    \centering
    \def\svgwidth{0.48\textwidth}
    \hspace*{-2pt}
    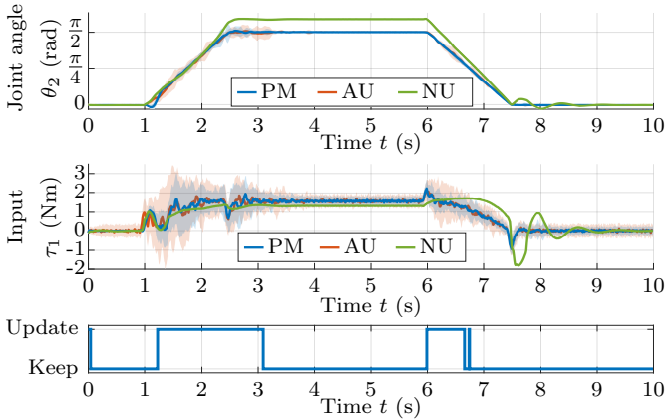
    \vspace*{-7mm}
    \caption{Mean trajectories and confidence intervals for Proposed Method, Always Update, and Never Update, and mean updating decisions of the proposed method.}
    \label{fig:traj}
\end{figure}

Fig.~\ref{fig:cost} shows boxplots of the total trajectory costs $J_{\text{tot}}= \sum_{k=0}^{1000} l\left(\gls{u}_{k},\, \gls{y}_{k}\right)$ for all methods. Never updating data (that is, continuously relying on the initial data) leads to a steady state error in the reference tracking (see  Fig.~\ref{fig:traj}) that results from the mismatch between predicted and actual system behavior. We observed a similar closed-loop behavior in a model-based setting, in which predictions are based on the linearization around the lower equilibrium (where the initial data is collected). Furthermore, Fig.~\ref{fig:cost} shows that always updating the data is not ideal either. Our proposed method yields a smaller variance of $J_{\text{tot}}$ and showed $18\,\%$ lower median cost in this simulation example.
The reason is that the proposed method does not require the addition of an exciting input signal $\gls{u}_k^{\text{rnd}}$. This influence can be further observed when investigating the mean trajectories and respective confidence intervals in Fig.~\ref{fig:traj}. For the shown ramp reference, both linear predictive controllers succeed in tracking with little error. However the proposed method exhibits oscillations smaller in amplitude for both inputs and outputs, as the inputs are not forced to be exciting. Note that only one input and one joint angle are shown for compact representation. The reference $\bm{y}^{\text{ref}}_{k}$ \linebreak was chosen to show steady-state behavior and not change too rapidly in between, since rapid reference changes lead to very large differences in the dynamics between prediction and real system, rendering the methods unstable. 

Fig.~\ref{fig:traj} also displays the mean updating decisions of the proposed method according to Algorithm~\ref{alg:method}.
As expected, the dataset is updated during transient behavior. During set point stabilization, new data are uninformative (i.e., would lead to violation of \eqref{eq:casehigh}) and thus rejected.

The SVD-based computation of the robustified rank in Algorithm~\ref{alg:method} is an order of magnitude faster than the mean computation time for solving \eqref{eq:ocp}, and thus does not add substantial computational load. Nevertheless, the performance of the proposed approach depends on the chosen threshold value $\rho$. For low threshold values, the dataset may be updated although new data are not sufficiently informative, yielding an invalid system representation.
Similarly for high threshold values, updates may be too infrequent, thus resulting in inaccurate predictions of the current system behavior, especially during transients.

\section{CONCLUSION} \label{sec:conclusion}
In this paper, we proposed an online data adaptation strategy for the approximate local representation of nonlinear systems from trajectory data. 
Based on the fundamental lemma and mosaic Hankel matrices, the data-driven system representation is linear and allows for the use of discontinuous input-output trajectories in its construction. 
In order to allow for an accurate behavioral approximation near the operating points, the data matrix is only updated if the data is sufficiently informative, i.e., the matrix does not violate the rank condition associated with persistency of excitation.
Since measurement noise artificially inflates the rank of the data matrix, we employ a robustified rank determined by the number of singular values above a certain threshold.
We applied the proposed data updating strategy in a simulation example of data-driven predictive control for a two-link robotic arm subject to measurement noise. The proposed approach showed improved control performance compared to both alternative strategies of using initially available data only, or updating the data in each time-step by relying on persistently exciting inputs.

Without injecting artificial excitation signals into the closed-loop input, the proposed method allows for a locally accurate data-driven system representation even near steady states and may serve in the design of linear direct data-driven controllers for nonlinear systems.

\def\bibfont{\small}
\bibliography{mybib}

\end{document}